# Quantum oscillations in field-induced correlated insulators of a moiré superlattice

Le Liu[1,2], Yanbang Chu[1,2], Guang Yang[1,2], Yalong Yuan[1,2], Fanfan Wu[1,2], Yiru Ji[1,2], Jinpeng Tian[1,2], Rong Yang[1,3], Kenji Watanabe[4], Takashi Taniguchi[5], Gen Long[3], Dongxia Shi[1,2,3], Jianpeng Liu[6,7], Jie Shen[1,2,3], Li Lu[1,2,3], Wei Yang[1,2,3*], and Guangyu Zhang[1,2,3*]

[1] *Beijing National Laboratory for Condensed Matter Physics and Institute of Physics, Chinese Academy of Sciences, Beijing 100190, China*
[2] *School of Physical Sciences, University of Chinese Academy of Sciences, Beijing, 100190, China*
[3] *Songshan Lake Materials Laboratory, Dongguan 523808, China*
[4] *Research Center for Functional Materials, National Institute for Materials Science, Tsukuba 305-0044, Japan*
[5] *International Center for Materials Nanoarchitectonics, National Institute for Materials Science, Tsukuba 305-0044, Japan*
[6] *School of Physical Sciences and Technology, ShanghaiTech University, Shanghai 200031, China*
[7] *ShanghaiTech Laboratory for Topological Physics, ShanghaiTech University, Shanghai 200031, China*

\* Corresponding authors. Email: wei.yang@iphy.ac.cn; gyzhang@iphy.ac.cn

**Abstract**

We report an observation of quantum oscillations (QOs) in the correlated insulators with valley anisotropy of twisted double bilayer graphene (TDBG). The anomalous QOs are best captured in the magneto resistivity oscillations of the insulators at $v = -2$, with a period of $1/B$ and an oscillation amplitude as high as 150 kΩ. The QOs can survive up to ~10 K, and above 12 K, the insulating behaviors are dominant. The QOs of the insulator are strongly $D$ dependent: the carrier density extracted from the $1/B$ periodicity decreases almost linearly with $D$ from $-0.7$ to $-1.1$ V/nm, suggesting a reduced Fermi surface; the effective mass from Lifshitz-Kosevich analysis depends nonlinearly on $D$, reaching a minimal value of 0.1 $m_e$ at $D = \sim -1.0$ V/nm. Similar observations of QOs are also found at $v = 2$, as well as in other devices without graphite gate. We interpret the $D$ sensitive QOs of the correlated insulators in the picture of band inversion. By reconstructing an inverted band model with the measured effective mass and Fermi surface, the density of state at the gap, calculated from thermal broadened Landau levels, agrees qualitatively with the observed QOs in the insulators. While more theoretical understandings are needed in the future to fully account for the anomalous QOs in this moiré system, our study suggests that TDBG is an excellent platform to discover exotic phases where correlation and topology are at play.

Quantum oscillations (QOs) of conductance are widely observed in mesoscopic devices[1, 2]. For metals in a magnetic field ($B$), QOs are usually revealed in the Shubnikov-de Haas oscillations (SdHOs) of conductance with a period of $1/B$ due to the Landau quantization of the Fermi surface; for insulators, however, SdHOs are in principle not expected, due to the absence of Fermi surface. Surprisingly, QOs have been observed in Kondo insulators[3-5] ($SmB_6$ and $YbB_{12}$), InAs/GaSb quantum wells[6, 7], and more recently in the insulating state of $WTe_2$[8]. To explain the anomalous QOs, various mechanisms are proposed, including the conventional picture with band inversion, in which the finite density of state (DOS) in the gap oscillates with $B$ due to the broadening of quantized Fermi surfaces away from gaps[9-14], unconventional Landau quantization of neutral Fermions[15-17], and trivial external capacitive mechanism due to the DOS in the gate electrode[18]. However, the strong debates impede the reaching of consensus. Noted that, despite the material differences, these insulators share one common feature, i.e. the insulating gaps are small and resulted from band hybridizations.

Twisted double bilayer graphene (TDBG), a highly tunable moiré flat band system that hosts correlated insulators[19-26], offers another opportunity to unveil the puzzling anomalous QOs. The bands in TDBG[27, 28] include flat moiré conduction bands and more dispersive moiré valence band and remote bands, allowing versatile band hybridizations. Besides, perpendicular electrical displacement field ($D$) could in-situ change the on-site energy between layers and in turn modify nonlinearly the moiré bandwidths as well as the gaps separating different bands[27, 29], enabling the facile tuning of Coulomb interactions. Moreover, there is a rich interplay of spin, valley, and Coulomb interaction that results in isospin competition with different polarized ground states of the correlated insulators at half fillings[30-32]. As demonstrated in TDBG[24], it could have a phase transition from metal or spin polarized correlated insulator to valley polarized correlated insulator by increasing $D$. Thus, TDBG provides a versatile playground for tuning electron correlation and band hybridizations. However, the QOs in the correlated insulators of TDBG as well as other twisted multilayer moiré systems[33-37] have been not explored.

Here, we report an observation of QOs in the correlated insulators of TDBG. We focus on the correlated insulator at $v = -2$ in the TDBG device with a graphite gate, and demonstrate the QOs in the insulator from the observation of the resistivity oscillations as a function of $1/B$. Then, we reveal that the insulating QOs are strongly $D$-field dependent, distinctly different from $D$ insensitive SdHOs of the Bloch electrons in the magnetic field. We also demonstrate the universality of the insulating QOs at $v = 2$ as well as in other devices without graphite gate. Lastly, we interpret the $D$ sensitive QOs of the correlated insulators in the picture of band inversion, and we could reproduce a finite DOS in the gap that oscillates with $1/B$ from the calculation of thermal broadened Landau levels (LLs).

The TDBG samples are prepared by using the 'cut and stack' method[38], and the details of device information are presented in our previous work[24]. These devices have a dual gate configuration (devices D1, D2 and D3 (D4) with a graphite (Si) bottom gate and gold top gate), which allows independent tuning of the carrier density $n$ and $D$. Here, $n = (C_{BG}V_{BG} + C_{TG}V_{TG})/e$ and $D = (C_{BG}V_{BG} - C_{TG}V_{TG})/2\varepsilon_0$, where $C_{BG}$ ($C_{TG}$) is the geometrical capacitance per area for bottom (top) gate, $e$ is the electron charge, and $\varepsilon_0$ is the vacuum permittivity. The filling factor is defined as $v = 4n/n_s$, corresponding to the number of carriers per moiré unit cell. Here, $n_s$ is the carrier density to fully fill a moiré band and defined as $n_s = 4/A \approx 8\theta/(\sqrt{3}a^2)$, where $A$ is the area of a moiré unit cell, $\theta$ is twisted angle, and $a$ is the lattice constant of graphene. In the following, we focus on the magneto transport of the correlated insulators at $v = -2$ in device D1 with twisted angle of 1.38°. All measurements are done at $T = 1.8$ K, unless stated otherwise.

Figure 1a-c shows a color mapping of longitudinal resistance $R_{xx}$ ($v$, $D$) at $B = 0$, 3 and 9 T, respectively. The correlated insulator at $v = -2$ develops at finite magnetic field due to orbital Zeeman effect[24]. It shows a periodic oscillating behavior with $D$ at $B = 3$ T in Fig. 1b. This pattern is similar to the LL crossings in layer-polarized systems[39, 40], where $D$ tunes the band alignment and further leads to conductance oscillations. Unlike the decoupled cases[39, 40] for metals near the band edge, however, the oscillations in Fig. 1b occurs in an insulator when moiré valence bands are half filled. Considering the valley degree of freedom in TDBG, there might be a complicated band structure with multiple Fermi surfaces at $v = -2$. At $B = 9$ T in Fig. 1c, the periodic oscillating behavior disappears and only the insulating state remains. To capture the evolution of band structure in magnetic field, we present a Landau fan diagram at $D^* = -0.94$ V/nm (Fig. 1d). LLs are observed fanning out from $v = -2$ with $v_{LL} = \pm 1, \pm 2, \pm 3, \pm 4, \pm 5, \pm 6$, indicating the presence of both electron-like and hole-like Fermi surfaces and a fully lifted degeneracy of spin and valley in the moiré valence band at such a high $D$. The interruption of the LLs at some finite $B$ indicates a change of LL degeneracy, which might due to LLs crossing from a complicated hybridized band structure. While these observations are in line with the previous report[24], however, the alternating color changes between red and white at $v = -2$ in Fig. 1d suggest resistance oscillations as a function of $B$.

The unexpected QOs are better presented in the plot of $R_{xx}$ ($B$) in Fig. 1e, with a nontrivial oscillation amplitude as high as 150 k$\Omega$ at around 6 T. The resistance oscillates periodically with $1/B$, as shown in Fig. 1f. By tracing the position of oscillation peaks and valleys, we obtain a frequency of $B_f \sim 13.3$ T. Here, the frequency can be related to a carrier density via $B_f = n\Phi_0/s$ with $s$ the degeneracy. These oscillations are very sensitive to the temperature, and they are barely visible at $T = 13$ K as shown in Fig. 2a. A full temperature dependence of the oscillations at $v = -2$ is depicted in Fig. 2b, where the negative and the positive sign of $dR/dT$ indicate insulator (blue) and metal (red), respectively. The metal insulator transition induced by orbital Zeeman effect occurs at $B \sim 1.2$ T (Fig. 2c) and then the QOs emerge in the insulating regime. Details of the insulating behavior for the QOs at $v = -2$ are revealed in the resistivity plots at different temperature in Fig. S1 and S2. In addition, evidence of the bulk insulating behavior at $v = -2$ is also revealed in the antiphase oscillations[7] of conductance $\sigma_{xx}$ and resistivity $\rho_{xx}$ (Fig. 2d) and $\sigma_{xy} \sim 0$ in correlated insulating regions (Fig. S3); for comparison, the conventional SdHOs from metallic states at $v = -0.32$ shows in phase oscillations (Fig. S4). Furthermore, the two terminal resistance transport measurements with edge grounded also suggests the QOs are originated from the insulating bulk states rather than edge states (Fig. S5). Thus, the observation of resistance oscillations at $v = -2$ is indeed a manifestation of QOs in insulator, distinct from SdHOs in metals.

The resistive QOs of the insulator at $v = -2$ are observed over a wide range of $|D| > 0.6$ V/nm, as shown in the color mapping of $R_{xx}$ ($D$, $B$) at $T = 100$ mK (Fig. 3a). In particular, these QOs emerge as a series of fans at $|D|$ from 0.7 to 1.1 V/nm, which is reminiscent of the Landau fan diagram with varied carrier density. By performing the fast Fourier transform (FFT) of Fig. 3a, the periodicity of these QOs, $B_f$, scales linearly to the change of $D$, indicated by the orange dashed line in Fig. 3b. Lastly, the QOs are vanishing at around $|D| \sim 1.17$ V/nm and then reentrant up on further increasing the $|D| > 1.2$ V/nm; however, the oscillating pattern at $|D| > 1.2$ V/nm is seemly antisymmetric compared to those at $|D| < 1.2$ V/nm, suggesting a pi phase change.

The observed $D$-field dependent resistive QOs at $v = -2$ are not due to the Brown-Zak oscillations[41] (BZOs, a feature of the Hofstadter butterfly). The BZOs are observed at a lower displacement field ($|D| < 0.6$ V/nm) in Fig. 3a with low resistance. Distinct from the QOs at $v = -2$ that strongly depends on $D$, the metallic BZOs frequency of $B_f = \sim 46$ T is independent on $D$ (right panel in Fig. 3b), corresponding to the commensuration of $\Phi/\Phi_0 = 1$ for moiré superlattices with a twisted angle of 1.38°. Such BZOs are also observed in the remote valance band ($v < -4$) at a high $|D|$ of 0.94 V/nm (Fig. S6), where the distinction between the resistive QOs at $v = -2$ and the metallic BZOs are demonstrated in their frequency difference. In addition, the insulating QOs at $v = -2$ are also not related to the crossing of LLs from $v = 0$ and that from $v = -4$. Since the latter means that the oscillation period could be traced back to a specific carrier density, i.e. a fixed frequency $B_f$, regardless of $D$. However, the observed $B_f$ at $v = -2$ is almost linearly dependent with $D$, which cannot be explained in the picture of LLs crossings from the CNP and the full filling.

Note that the QOs of the insulator at $v = -2$ have also been observed in other three devices, D2 of $\theta = 1.21°$ (Fig. S7) and D3 of $\theta = 1.51°$ (Fig. S8) with a metal top gate and a graphite back gate, and D4 with $\theta = 1.26°$ (Fig. S9) with a metal top gate and a heavy doped Si back gate. Similar to D1, the resistive QOs and the $D$ tunable low oscillation frequency ($B_f < 25$ T) are different from the BZOs. The observations in both graphite and non-graphite gate devices indicate that the resistive QOs cannot be induced by capacitive mechanism[18] of graphite DOS at high magnetic fields. Besides, we observe the similar QOs of the insulator at $v = 2$ in Fig. S10. These observations indicate that the QOs at half fillings are universal for correlated insulators with valley anisotropy where a general model might apply.

In the following, we interpret the insulating QO at $v = -2$ in the picture of band inversion[9-14]. The center idea is to consider the thermal broadening of the LLs from inverted bands, and it will predict a finite DOS in the gap that oscillates with $1/B$, with the $B_f$ defined by the enclosed area of inverted bands without hybridization. Here, the assumption of band inversion at the correlated insulator is motivated from the single-particle band structure calculation in Fig. S11a and Hartree-Fock calculation of the ground states in TDBG[24]. In continuum model, the finite $D$ breaks C2 symmetry, which leads to the inequivalence of k and k' points in moiré Brillouin zone. The two bands from K and K' valley constitute an inverted band at these two points due to the time reversal symmetry. Furthermore, this inverted band could evolve from overlapped bands to gapped bands with the hybridization between two valley subbands (partially valley polarized with intervalley coherence). The potential energy difference between two valley subbands K and K' is the sum of the orbital Zeeman effect and Coulomb interaction, described by the formula $\Delta E = 2g\mu_B B + E_c$, where $g$, $\mu_B$, and $E_c$ are effective g factor, Bohr magneton, and Coulomb interaction energy, respectively. Since the QOs onset at a low magnetic field of ~1.5 T, and the corresponding Zeeman splitting energy should be much smaller than the bandwidth of the two subbands. We further assume that the overlap of the two subbands are determined by $E_c$. This assumption is supported by the fact that the observed $B_f$ from QOs at $v = -2$ is independent of $B$. Note that the overlap is linearly dependent on $D$ in Fig. 3b. The observation indicates that $E_c$ scales linearly with $D$, in agreement with the $D$ dependent bandwidth of the moiré valance in Fig. S11b.

By doing quantitively analysis, we obtain effective mass $m^*$ and oscillating carrier density $n$ from QOs of the insulator. As shown in Fig. 3c, similar to the SdHOs in metals, the temperature dependence of QOs in the insulator are found following the Lifshitz–Kosevich formula: $\Delta R = R_{xx}(T) - R_{xx}(T = 13\text{ K}) \sim kT/\sinh(kT)$, where $k = 2\pi^2 k_B m^*/(\hbar eB)$. Here, $k_B$, $m^*$, $\hbar$, and $e$ are Boltzmann constant, effective mass, reduced Planck constant, and electron charge, respectively. The obtained $m^*$ from QOs of the insulator varies from 0.1 to 0.15 $m_e$ ($m_e$ is electron mass) in the oscillation region, decreasing with the increase of $|D|$ at first and then rising rapidly after reaching the minimum value close to $D = -1$ V/nm, as shown in Fig. 3d. In the same way, we obtain $m^*$ in Fig. S12 for both the SdHOs and the insulating QOs at $D = -0.94$ V/nm. In addition, the extracted oscillating carrier density $n$ is shown as an orange line in Fig. 3d, via $B_f = A\Phi_0/(2\pi)^2 = n\Phi_0/s$, where $A$ is the enclosed area of inverted bands without hybridization[11] and $s = 2$ is the spin degeneracy of valley subbands.

Then, we calculate the LLs spectra of the inverted bands at $\nu = -2$, by considering the thermal broadening and the experimental values of gap size $\Delta$, $m^*$, and oscillating carrier density $n$. Fig. 4a shows a reconstructed band structure of the moiré valence band at $D = -0.94$ V/nm, with $m_K = 0.15$ $m_e$ and $m_{K'} = m_K/0.8$, $\Delta \sim 1.6$ meV, $n \sim 6.5 \times 10^{11}$ /cm$^2$. The resulted LLs spectrum with inverted bands at $T = 2$ K are shown in top panel of Fig. 4b. The finite low energy DOS in the gap oscillate with $1/B$, which agrees with our experimental data in bottom panel of Fig. 4b. Moreover, the model establishes a close relationship between the size of the enclosed Fermi surface and $D$, as the latter governs the size of band overlap $\delta\mu$ (details in Supplementary Note 5). The resulted calculations in Fig. 4c indeed predict a change of QOs periodicity as a function of $|D|$, in good agreement with our observation in Fig. 3a. Last but not least, we further carry out numerical calculations to show the LL spectra away from $\nu = -2$, as shown in the comparison between experimental data of longitudinal conductivity and the calculated DOS in Fig. S13. The inverted band model predicts LL intersections away from $\nu = -2$, which qualitatively matches pretty well with experimental crossing points, indicated by the red circles. The agreement between experimental data and the theoretical model further supports our interpretation.

While the model can well explain the data at $|D|$ from 0.7 to 1.1 V/nm, it fails to account for the observations at $|D| \geq \sim 1.2$ V/nm. This discrepancy lies in the over-simplified inverted band model and might suggest a more complicated band structure when Coulomb interaction plays an important role. Also noted, the real systems might acquire phases as $D$ changes, as shown in Fig. S14; however, phase changes are not included in the inverted band model. Besides, the QOs of insulators are supposed to disappear at ultra-low temperature due to the negligible thermal activation, while we observed robust QOs even at $T = 100$ mK (Fig. 3a and Fig. S15). The saturated oscillation amplitude in zero temperature limit is probably ascribed to the lifetime broadening of LLs[13] or formation of excitonic insulating behavior[14].

Lastly, we briefly discuss the other possible interpretation of the neutral Fermi surfaces[15-17]. The correlated insulator with valley anisotropy at half fillings in TDBG, together with trigonal moiré superlattice and spin unpolarized, is a possible system with frustrated magnetic interactions for hosting quantum spin liquid states[42, 43]. In this exotic picture, the total resistance is the sum of bosonic charge insulator (non-oscillating) and fermionic spinon (oscillating). To testify the neutral Fermion picture, however, conventional charge transport in our work and others falls short, and experimental techniques that could distinguish the charge transport and spin transport are highly demanded.

To sum up, we have observed anomalous oscillating behaviors in the correlated insulator at $v = -2$ in the TDBG, with a period of $1/B$ and an oscillation amplitude as high as 150 k$\Omega$. Such insulating QOs are strongly tunable with respect to the $D$-field, distinctly different from $D$ insensitive SdHOs of the Bloch electrons in the magnetic field. We have also demonstrated the universality of the insulating QOs at $v = 2$ as well as in other devices with different twist angles and gate configurations. The observations are captured in the picture of band inversion, which reproduces a finite DOS in the gap that oscillates with $1/B$ from the calculation of thermal broadened LLs. Being an electric field tunable system, TDBG is an excellent platform to study the oscillations in an insulator. The evolution of valley subbands from band inversion to complete separation suggests abundant correlation-driven behaviors and emerging topological phases in field-tunable TDBG systems. While most of the QOs in the correlated insulators are captured in the phenomenological model of the inverted bands with hybridizations, more theoretical and experimental investigations are needed in the future to better understand the insulating QOs in TDBG and other moiré systems as well.

## Acknowledgements

We thank Gang Li, Kun Jiang, Zhida Song, Yuan Wan, Quansheng Wu for useful discussions. We acknowledge supports from the National Key Research and Development Program (Grant No. 2020YFA0309600), National Science Foundation of China (NSFC, Grant Nos. 61888102, 11834017, 1207441), and the Strategic Priority Research Program of CAS (Grant Nos. XDB30000000& XDB33000000). Measurements at sub-Kelvin temperature are supported by the Synergetic Extreme Condition User Facility at CAS. Gen Long acknowledge the support from NSFC (Grant No. 12104330). Kenji Watanabe and Takashi Taniguchi acknowledge support from the Elemental Strategy Initiative conducted by the MEXT, Japan (Grant No. JPMXP0112101001), JSPS KAKENHI (Grant Nos. 19H05790, 20H00354 and 21H05233) and A3 Foresight by JSPS.

## Author contributions

Wei Yang and Guangyu Zhang supervised the project; Wei Yang and Le Liu designed the experiments; Le Liu and Yanbang Chu fabricated the devices and performed the magneto-transport measurement with the assistance of Guang Yang, Jie Shen, Li Lu, Yalong Yuan, FanfanWu, Yiru Ji, Jinpeng Tian, Rong Yang, Gen Long and Dongxia Shi; Le Liu and Jianpeng Liu calculated the band structure with a continuum model and Landau level DOS in an inverted band model; Kenji Watanabe and Takashi Taniguchi provided hexagonal boron nitride crystals; Wei Yang, Le Liu, and Guangyu Zhang analyzed and interpreted the data; Wei Yang and Le Liu wrote the paper with the input from all the authors.

## Data availability

The data that support the findings of this study are available from the corresponding authors upon reasonable request.

## Competing interests

The authors declare no competing interests.

## Additional information

Supplementary information is provided online.

# Figure & Figure captions

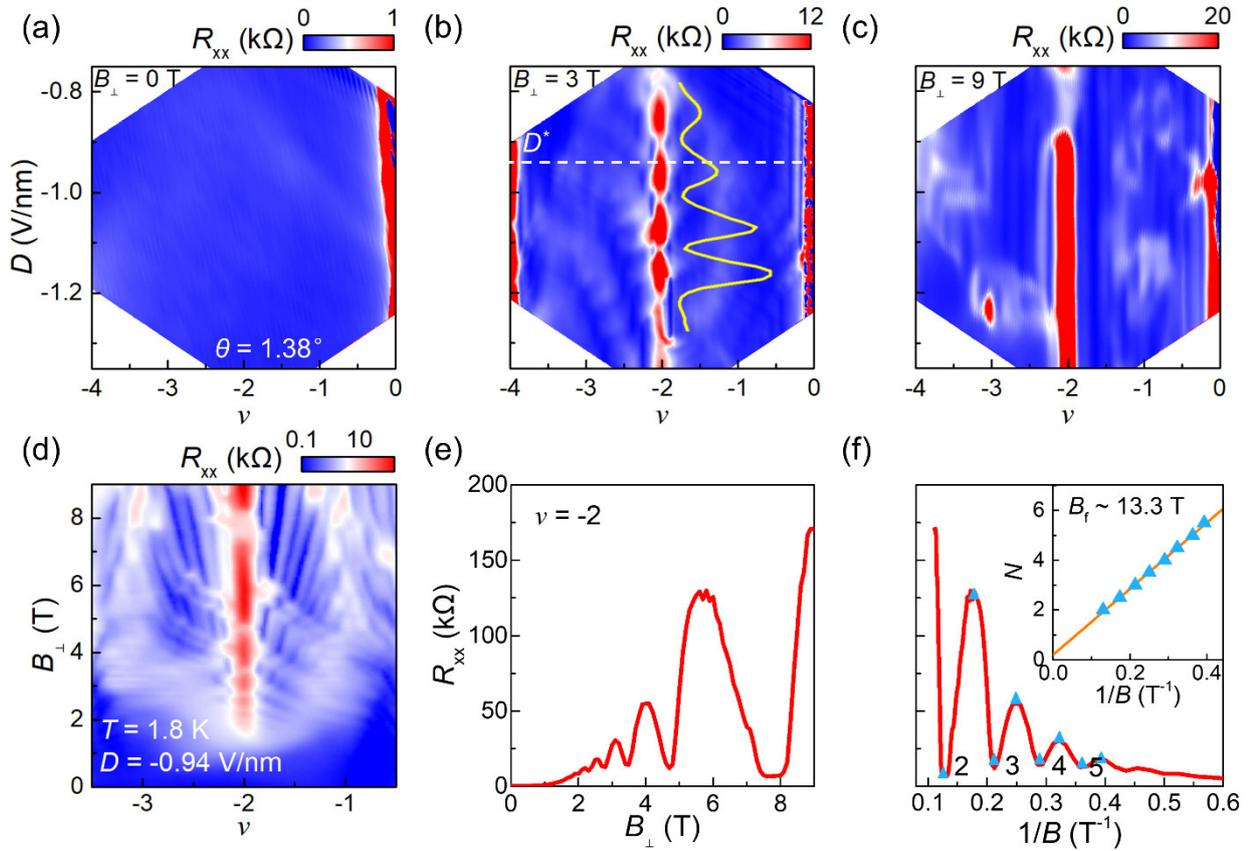

**Fig. 1.** Quantum oscillations of $v = -2$ insulators. (a-c) $R_{xx}$ as a function of $v$ and $D$ at $B_\perp = 0$, 3 and 9 T, respectively. The yellow line in (b) is a line cut along $v = -2$. (d) Landau fan diagram at $D = -0.94$ V/nm and $T = 1.8$ K. (e) Line cuts at $v = -2$ shown in (d). (f) $R_{xx}$ versus $1/B$ at $v = -2$. Inset, oscillation index versus $1/B$. Blue triangles correspond to the position of oscillation peak and valley. The Orange line corresponds to the linear fitting.

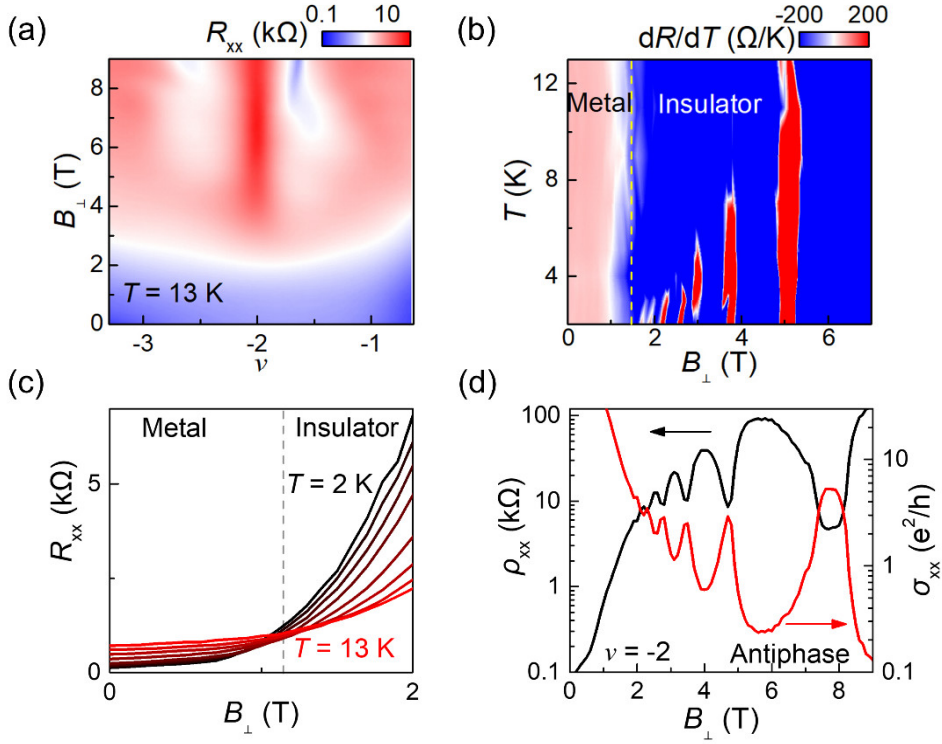

**Fig. 2.** Temperature-dependent quantum oscillations. (a) Landau fan diagram at $D = -0.94$ V/nm and $T = 13$ K. (b) First order derivative of $R_{xx}$ ($T$) as a function of $B_\perp$ and $T$ at $v = -2$. (c) Linecuts of $R_{xx}$ ($B$) at different temperature. (d) $\rho_{xx}$ and $\sigma_{xx}$ as a function of $B_\perp$ at $v = -2$.

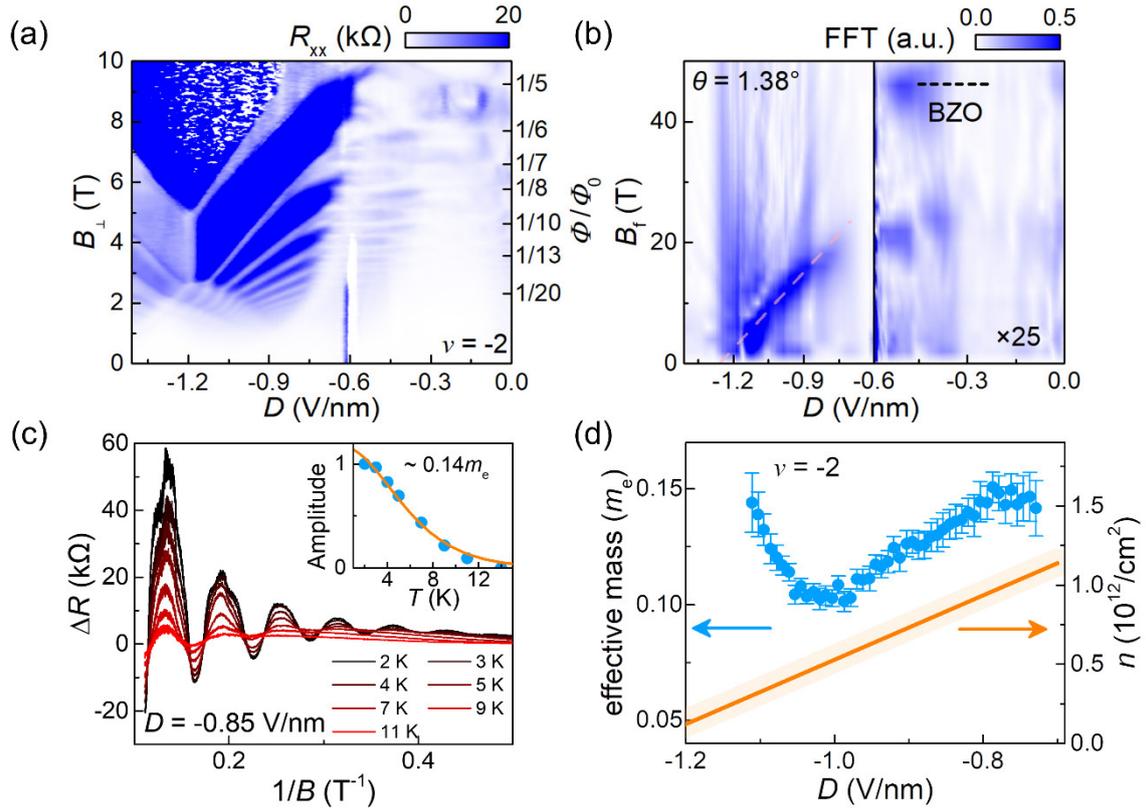

**Fig. 3.** Electric field tunable quantum oscillations. (a) $R_{xx}$ as a function of $D$ and $B_\perp$ at $v = -2$ and $T = 100$ mK. (b) FFT of (a). Amplitude is magnified 25 times in a range of $D = 0$ to $-0.6$ V/nm. the orange dash line is a linear fitting along the amplitude peak. (c) $\Delta R$ versus $1/B$ at different temperature. Inset, L-K fitting of the oscillation amplitude. (d) Effective mass and carrier density versus $D$ at $v = -2$. Error bars of effective mass are estimated from the least square method. The orange shadow corresponds to the uncertainty of carrier density calculated from FFT results.

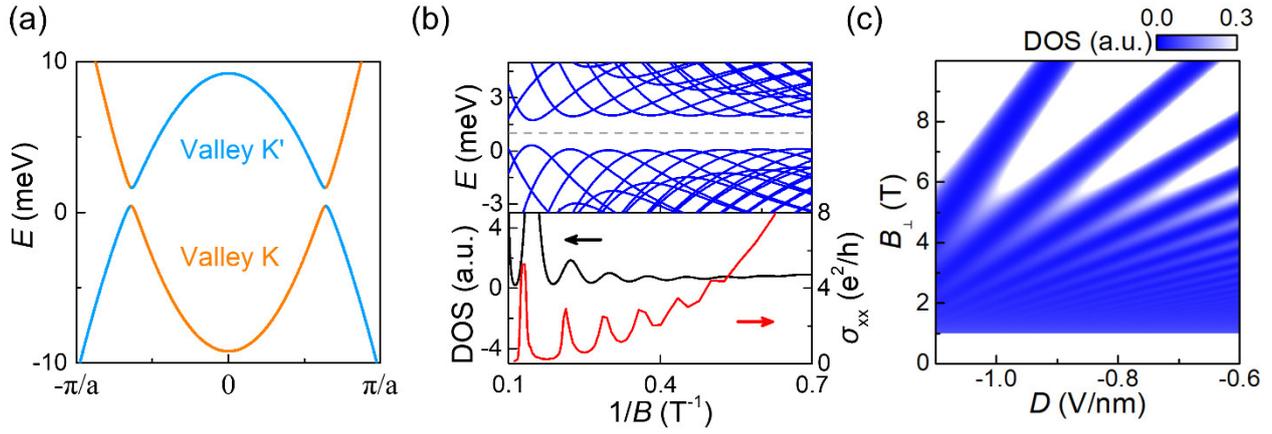

**Fig. 4.** Oscillation in hybridization gaps of inverted band systems. (a) Hybridized inverted Band model. The orange (blue) line corresponds to the energy band in valley K(K'). (b) Top, calculated Landau levels spectrum with $1/B$. The gray line corresponds to the chemical potential. Bottom, calculated low energy DOS in gap (black line) and measured $\sigma_{xx}$ at $v = -2$ and $D = -0.94$ V/nm (red line). (c) Calculated DOS spectrum as a function of $D$ and $B_\perp$.

# Supplementary Materials for
# "Quantum oscillations in field-induced correlated insulators of a moiré superlattice"

Le Liu[1,2], Yanbang Chu[1,2], Guang Yang[1,2], Yalong Yuan[1,2], Fanfan Wu[1,2], Yiru Ji[1,2], Jinpeng Tian[1,2], Rong Yang[1,3], Kenji Watanabe[4], Takashi Taniguchi[5], Gen Long[3], Dongxia Shi[1,2,3], Jianpeng Liu[6,7], Jie Shen[1,2,3], Li Lu[1,2,3], Wei Yang[1,2,3*], and Guangyu Zhang[1,2,3*]

[1] *Beijing National Laboratory for Condensed Matter Physics and Institute of Physics, Chinese Academy of Sciences, Beijing 100190, China*

[2] *School of Physical Sciences, University of Chinese Academy of Sciences, Beijing, 100190, China*

[3] *Songshan Lake Materials Laboratory, Dongguan 523808, China*

[4] *Research Center for Functional Materials, National Institute for Materials Science, 1-1 Namiki, Tsukuba 305-0044, Japan*

[5] *International Center for Materials Nanoarchitectonics, National Institute for Materials Science, 1-1 Namiki, Tsukuba 305-0044, Japan*

[6] *School of Physical Sciences and Technology, ShanghaiTech University, Shanghai 200031, China*

[7] *ShanghaiTech Laboratory for Topological Physics, ShanghaiTech University, Shanghai 200031, China*

\* Corresponding authors. Email: wei.yang@iphy.ac.cn; gyzhang@iphy.ac.cn

**Supplementary Note 1: Transport measurements**

The low temperature magneto-transport measurements are performed in a helium-4 cryostat (base temperature ~ 1.8K) and an Oxford dilution refrigerator (base temperature ~ 10mK). We measure the four terminal longitudinal resistance $R_{xx}$ and Hall resistance $R_{xy}$ using the lock-in amplifier SR830/LI5650 with a frequency of 10~35Hz. Excitation currents $I$ = 10nA with a large series resistance 100MΩ or excitation voltages $V$ = 200uV with a 1/1000 divider are used in our measurements.

To give more information about the insulator at $v$ = -2, we plot $\rho_{xx}$ versus $B$ or $T$ in Fig. S1 by eliminating the geometry factors of device D1 (aspect ratio is 1.4). As shown in Fig. S1(a) and 1(b), the metal-insulator transition occurs at $B$ ~1.2T, followed by the quantum oscillations. The temperature dependence of the resistance peak shows the thermal activation behaviors, which is the sign of an insulator. Besides, the temperature dependence of the resistance dip shows a non-monotonic behavior, increasing like a metal at first and then decreasing like an insulator. This behavior suggests the insulating nature at $v$ = -2 when the quantum oscillations are smeared by thermal effect. As shown in Fig. S2, we extract the energy gaps according to Arrhenius thermal activation formula. The energy gap appears when B > 1.2T and grows with the increase of magnetic field, up to 1.5meV at B = 2.5T. Furthermore, as shown in the $\sigma_{xy}$ maps (Fig. S3), $\sigma_{xy}$ ~ 0 in the region with QOs also suggests an insulating nature. The oscillations of $\rho_{xx}$ and $\sigma_{xx}$ show antiphase behaviors at $v$ = -2 while in phase behaviors at $v$ = -0.32. Here $\sigma_{xx} = \rho_{xx}/(\rho_{xx}^2 + \rho_{xy}^2)$. The latter is according with SdH oscillations in 2D metal due to the vanishing $\rho_{xx}$ and large $\rho_{xy}$. However, $\rho_{xx}$ is much larger than $\rho_{xy}$ in an insulator. Hence, we expect to observe the antiphase oscillations[1] as shown in Fig. S4.

We have carried out transport measurements on device D3 with twisted angle of 1.51° using different measurement configurations to distinguish the contribution from edge and bulk. The schematics of circuits are presented in Fig. S5(a-b), where the left circuit excludes any contribution from the edge current while the right includes it. As shown in Fig. S5(c-d), the two kinds of measurements exhibit different behaviors for SdHOs at $v = -0.3$ and QOs at $v = -2$. For the former, the bulk resistance $R_{2t\_Bulk}$ and longitudinal resistance $R_{xx}$ exhibit antiphase oscillations. In other words, the minimum $R_{xx}$ of edge transport corresponds to the maximum $R_{2t\_Bulk}$ of bulk transport, in accordance with the coexistence of insulating bulk states and conducting edge states in the Landau quantization region. However, for the latter, the resistance from two kinds of measurements shows in-phase oscillations. This behavior suggests the QOs at $v = -2$ are originated from the hybridized gap of bulk without any edge states.

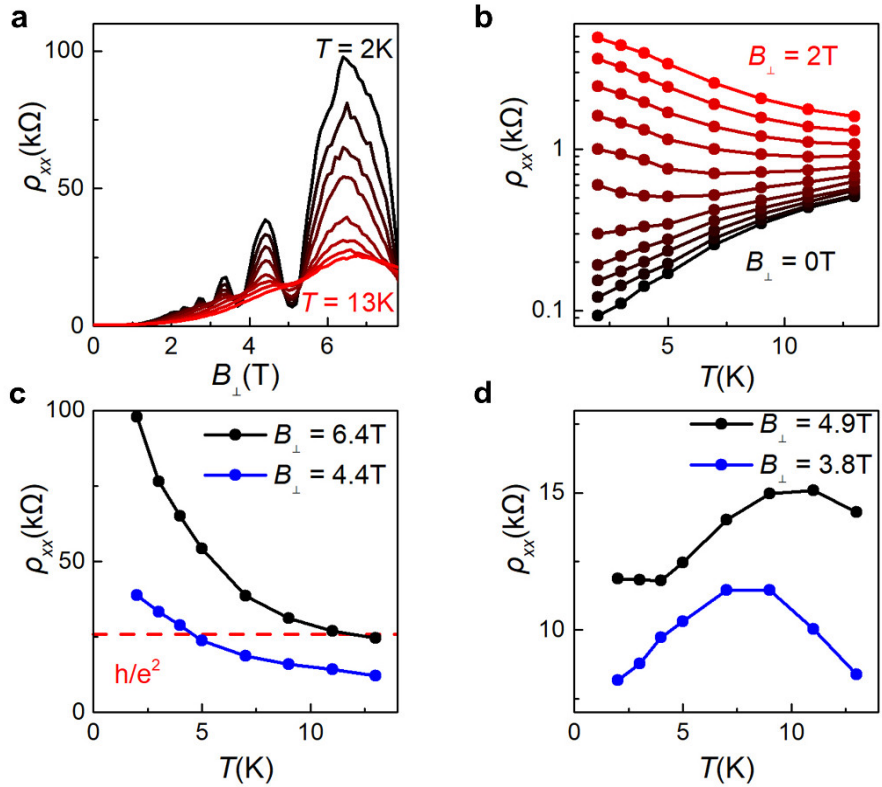

**Fig. S1. Evidence of insulating behaviors at $v = -2$.** **a** Temperature-dependent quantum oscillations at $v = -2$ and $D = -0.94$V/nm. **b** Metal-insulator transitions triggered by magnetic field at $v = -2$. **c, d** shows the temperature-dependent resistivity at peaks and dips of the oscillations, suggesting the universal insulating behaviors at $T > 10$K.

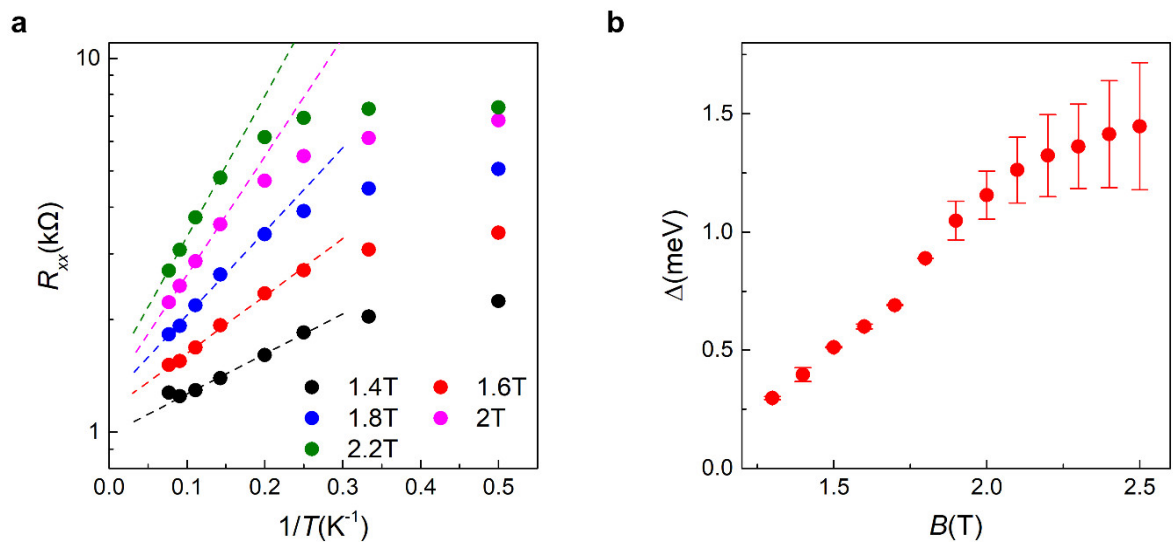

**Fig. S2. Thermal activated energy gap at $v = -2$ near the region of metal-insulator transition. a** Using Arrhenius plots to extract the energy gap. **b** Energy gaps versus magnetic field.

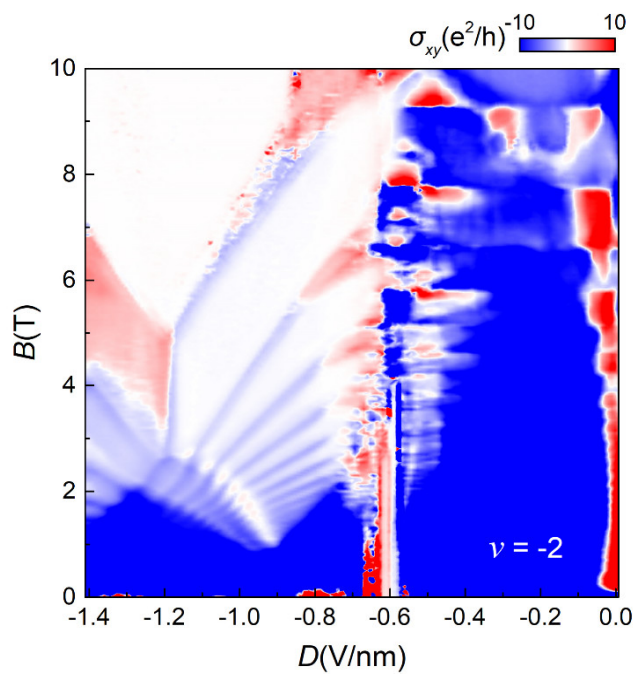

**Fig. S3.** $\sigma_{xy}$ **maps as a function of $D$ and $B$ at $v = -2$.**

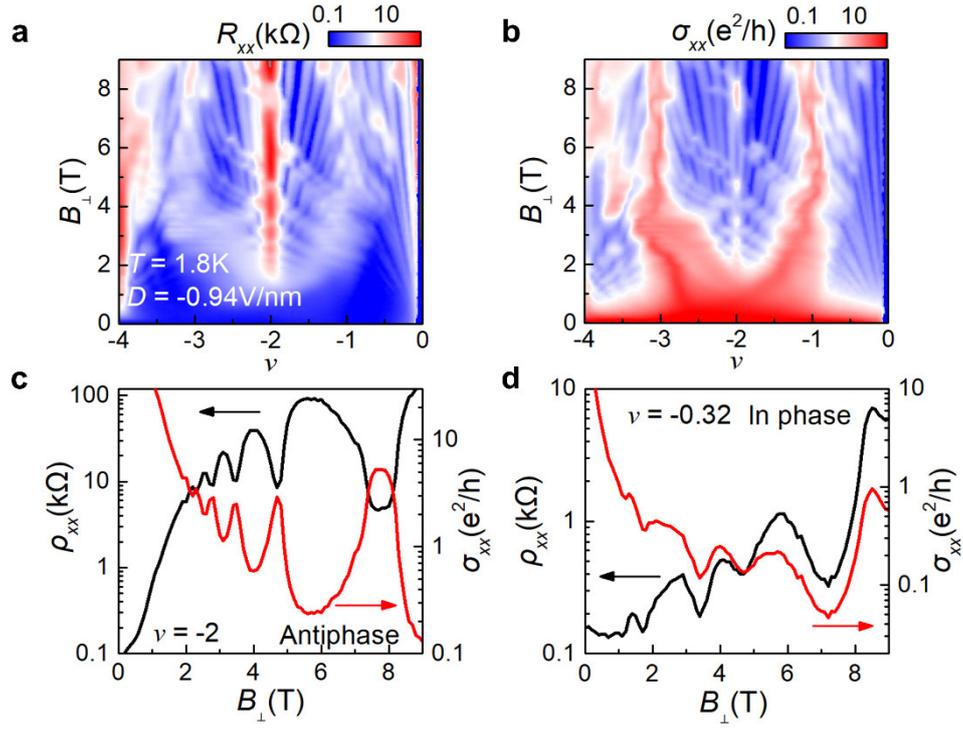

**Fig. S4. Antiphase oscillations at $v = -2$. a, b** $R_{xx}$ and $\sigma_{xx}$ as a function of $v$ and $B_\perp$ at $D = -0.94$V/nm. **c, d** $\rho_{xx}$ and $\sigma_{xx}$ as a function of $B_\perp$ at $v = -2$ and $v = -0.32$, respectively.

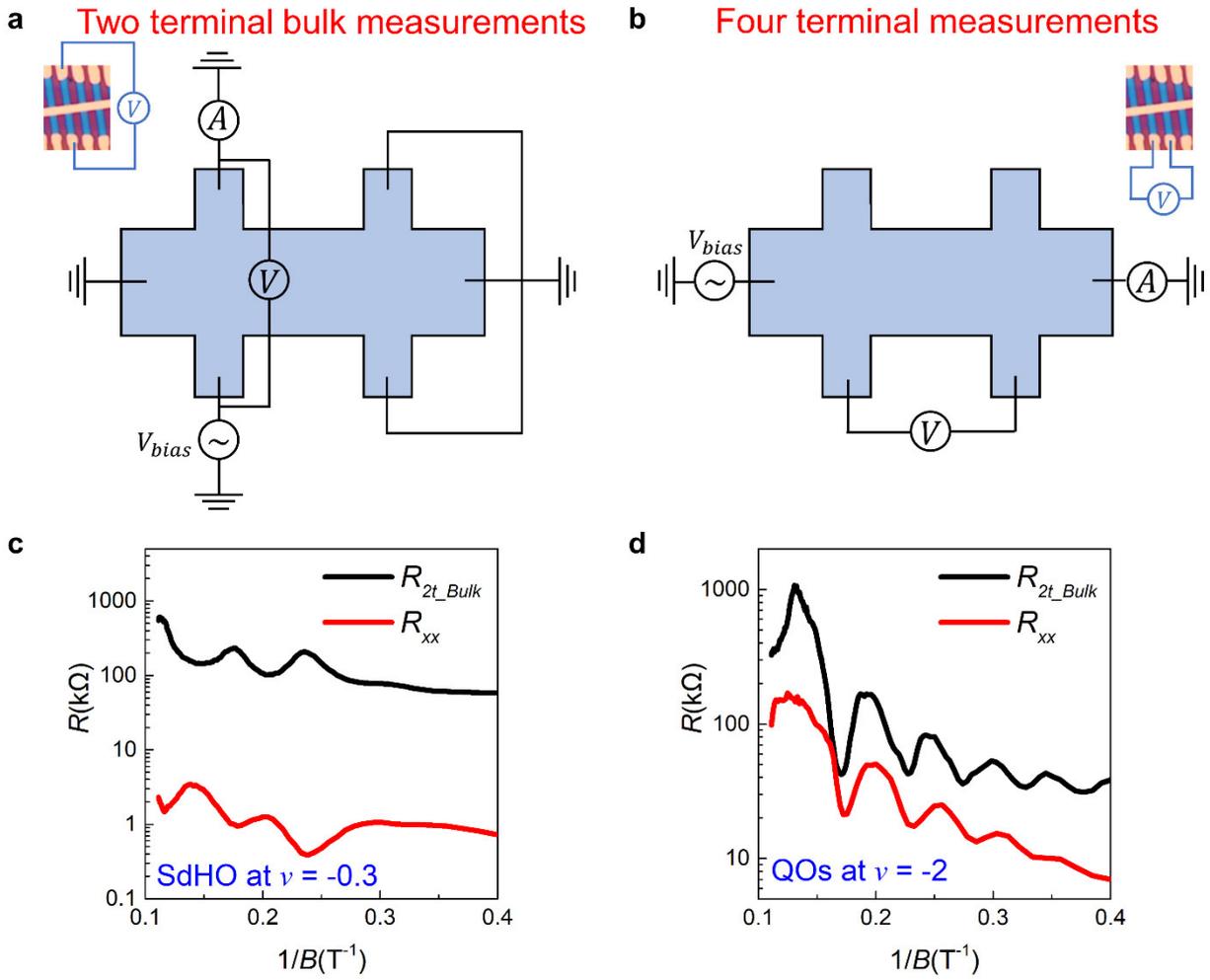

**Fig. S5. Different measurement configurations. a** two terminal resistance measurement with edges grounded, to exclude the edge state contribution. **b** Typical four terminal Hall bar measurements. The inset is the optical image of device D3. **c** $R_{2t\_Bulk}$ and $R_{xx}$ versus $1/B$ at $v = -0.3$ and $D = -1.2$V/nm. **d** $R_{2t\_Bulk}$ and $R_{xx}$ versus $1/B$ at $v = -2$ and $D = -1.2$V/nm. The resistance difference lies in the channel length difference between 2 terminal measurement and the 4 terminal Hall bar measurement.

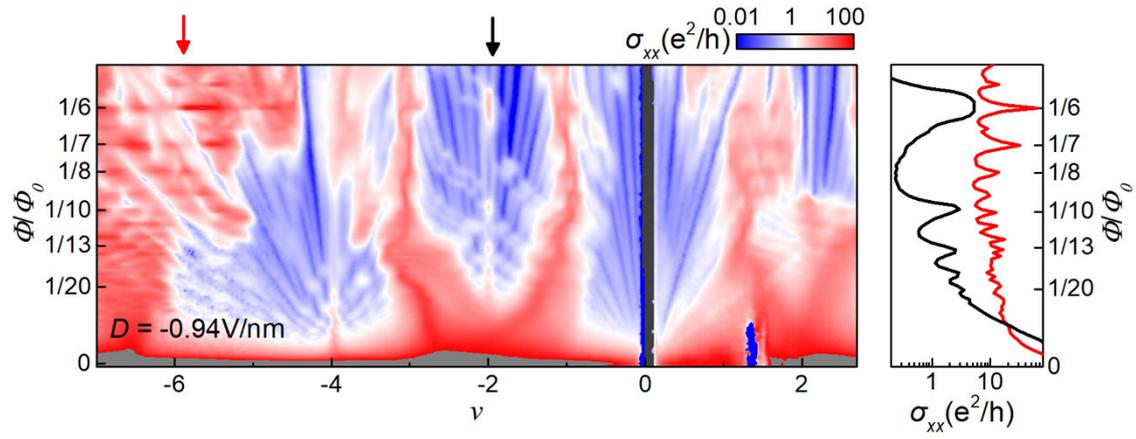

**Fig. S6. Wannier diagram at D = -0.94V/nm.** The red arrow corresponds to Brown-Zak oscillation[2]. The black arrow corresponds to quantum oscillations of insulators. Right figure, line cuts of the two types of oscillations shown in left figure.

## Supplementary Note 2: Quantum oscillations of an insulator in other devices

We replicate our observations of the quantum oscillations at $v = -2$ in other three devices. The first (D2) with twisted angle of 1.21° has Ti/Au top gate and graphite bottom gate; the second (D3) with twisted angle of 1.51° has Ti/Au top gate and graphite bottom gate; the third (D4) with twisted angle of 1.26° has Ti/Au top gate and Si(p++) bottom gate.

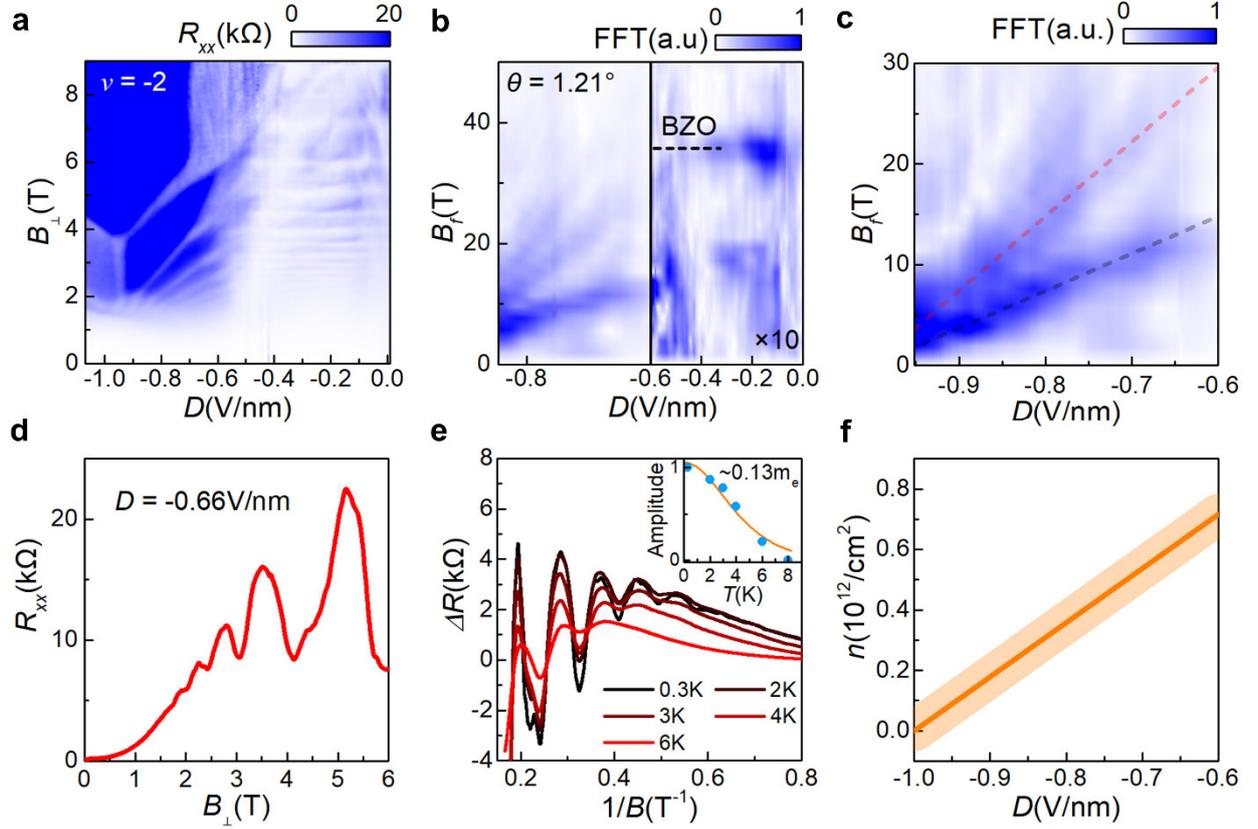

**Fig. S7. Quantum oscillations of $v = -2$ insulators in device D2. a** $R_{xx}$ as a function of $D$ and $B_\perp$ at $v = -2$ and $T = 100$mK. **b** Fast Fourier transform (FFT) of **a**. Amplitude is magnified 10 times in a range of $D = 0$ to $-0.6$V/nm. **c** Magnified figure of **b**. Black and red lines correspond to the linear fitting of the first and second order FFT peaks of quantum oscillations, respectively. **d** Line cuts at $D = -0.66$V/nm shown in **a**. **e** $\Delta R_{xx}$ versus $1/B$ at different temperature. Here $\Delta R = R(T) - R(T = 8K)$. Inset, L-K fitting of the oscillation amplitude. **f** Carrier density versus $D$ at $v = -2$. The orange shadow corresponds to the uncertainty of carrier density calculated from FFT maps.

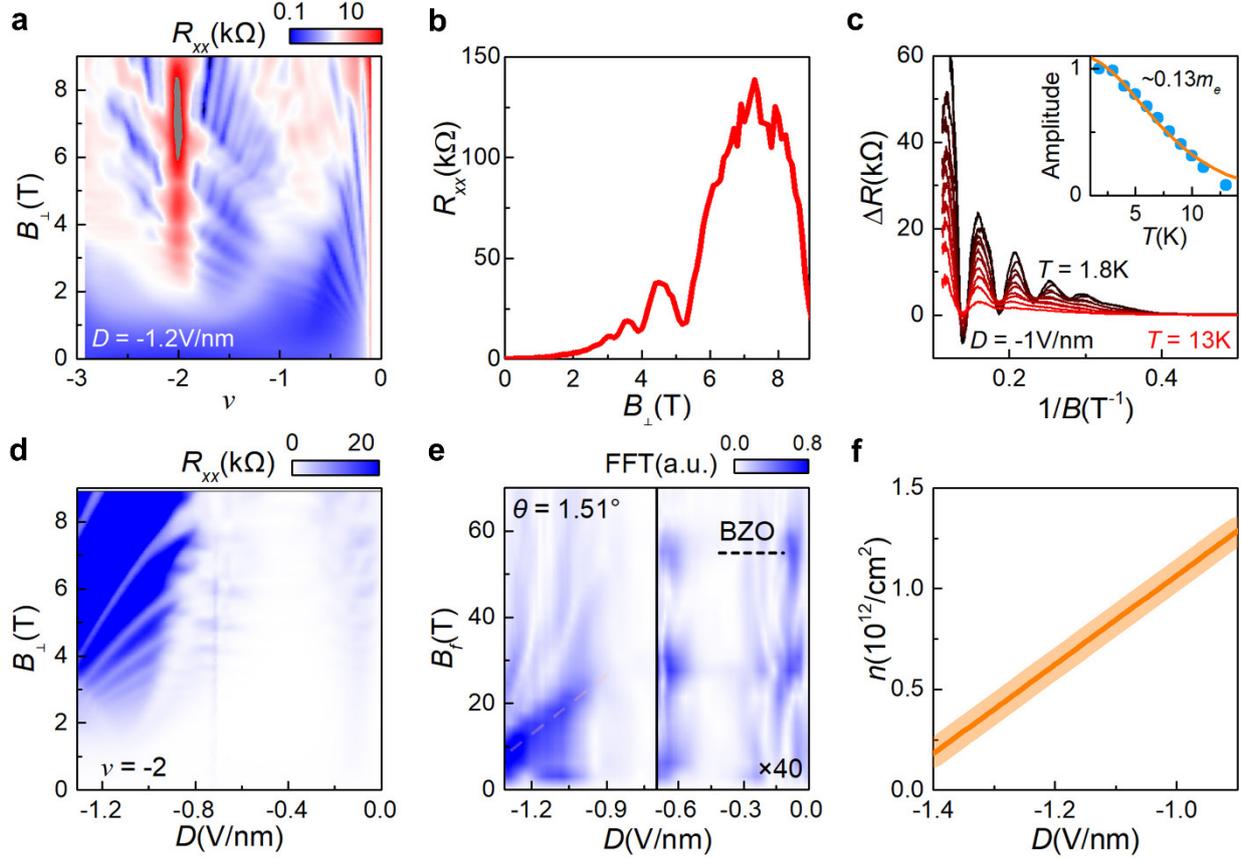

**Fig. S8. Quantum oscillations of $v = -2$ insulators in device D3. a** $R_{xx}$ as a function of $v$ and $B_\perp$ at $D = -1.2$V/nm and $T = 1.8$K. **b** Line cuts at $v = -2$ shown in **a**. **c** $\Delta R_{xx}$ versus $1/B$ at different temperature. Here $\Delta R = R(T) - R(T = 15$K$)$. Inset, L-K fitting of the oscillation amplitude. **d** $R_{xx}$ as a function of $D$ and $B_\perp$ at $v = -2$. **e.** Fast Fourier transform (FFT) of **d**. Amplitude is magnified 40 times in a range of $D = 0$ to $-0.7$V/nm. **f** Carrier density versus $D$ at $v = -2$. The orange shadow corresponds to the uncertainty of carrier density calculated from FFT maps.

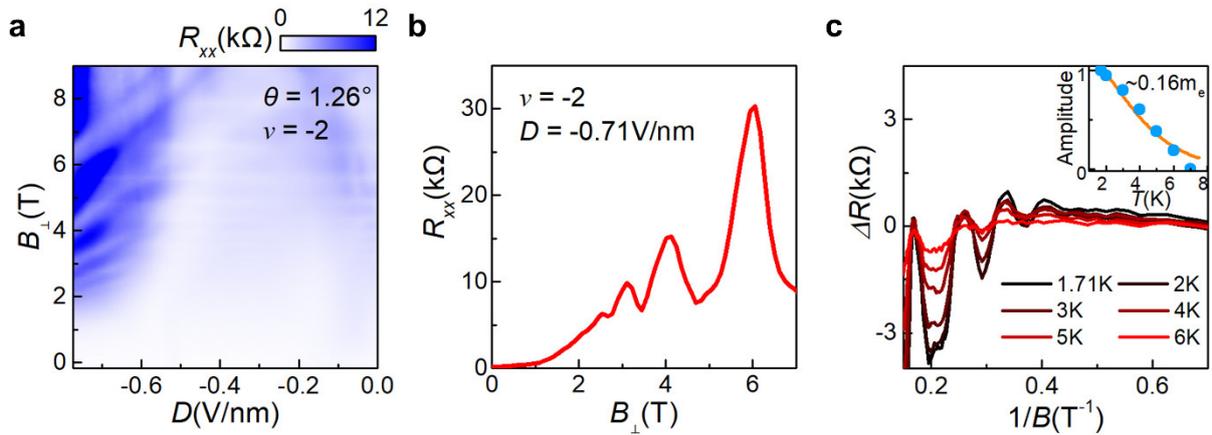

**Fig. S9. Quantum oscillations of $v = -2$ insulators in device D4. a** $R_{xx}$ as a function of $D$ and $B_\perp$ at $v = -2$ and $T = 1.8$K. **b** Line cuts at $D = -0.71$V/nm shown in **a**. **c** $\Delta R_{xx}$ versus $1/B$ at different temperature. Here $\Delta R = R(T) - R(T = 7$K$)$. Inset, L-K fitting of the oscillation amplitude.

## Supplementary Note 3: Quantum oscillations of $v = 2$ insulators.

As shown in Fig. S10(a), the state at $v = 2$ evolves from a metal to a spin-polarized insulator at the modest $D$ from ~0.3 to ~0.6V/nm. It then becomes a valley-polarized insulator when $D$ is larger than 0.6V/nm. Similar to the quantum oscillations at $v = -2$, the state also oscillates with $B_\perp$ and the behaviors are shown in Fig. S10(b) and 10(c).

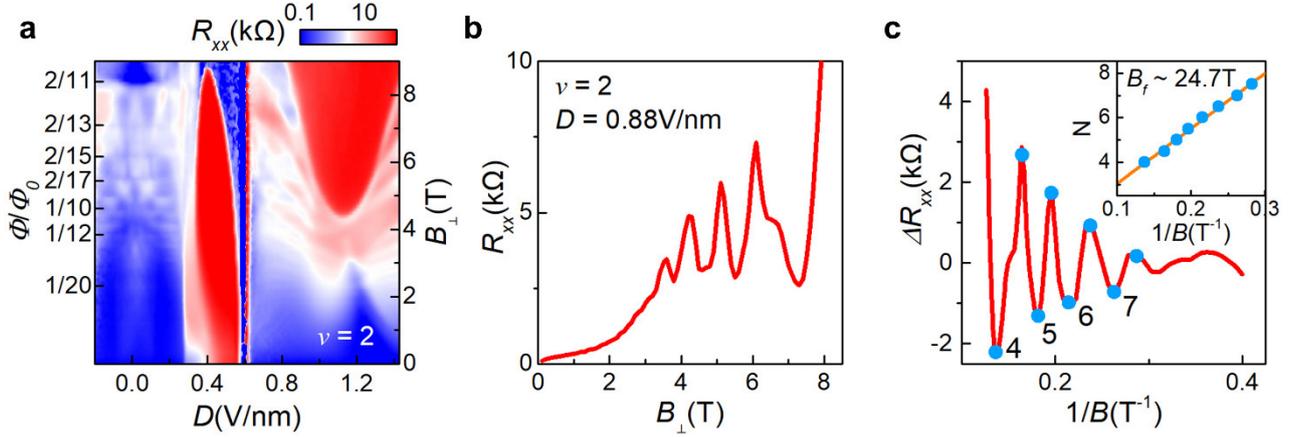

**Fig. S10. Quantum oscillations at $v = 2$. a**, $R_{xx}$ as a function of $D$ and $\Phi/\Phi_0$ at $v = 2$. **b**, Line cuts at $D = 0.88$V/nm shown in **a**. **c**, $\Delta R_{xx}$ versus $1/B$. Here the smooth polynomial background is subtracted. Inset, oscillation index versus $1/B$.

## Supplementary Note 4: Band structure calculations

We calculate the band structure of AB-BA stacked TDBG with a twisted angle $\theta = 1.38°$ using the continuum model[3, 4]. The interlayer coupling terms are set as $u_{AA} = 80$meV, $u_{AB} = 100$meV (A, B correspond to sublattice indices) due to the relaxation effect. Electric field tunable interlayer asymmetry potential is

$$V = \begin{pmatrix} \frac{3}{2}U\hat{1} & 0 & 0 & 0 \\ 0 & \frac{1}{2}U\hat{1} & 0 & 0 \\ 0 & 0 & -\frac{1}{2}U\hat{1} & 0 \\ 0 & 0 & 0 & -\frac{3}{2}U\hat{1} \end{pmatrix},$$

where U is the interlayer potential energy difference. The calculated band structures at different U are shown in Fig. S11(a). The bandwidth of conduction band and valance band can be extracted from the band structure calculation (Fig. S11(b)). It decreases with U from U=40 to 80meV, coincident with the quantum oscillation region in experiments from $D = -0.6$ to $-1.2$V/nm. Fig. S11(c) shows the density of states (DOS) of valance band at different U. The van Hove singularity with diverge DOS gradually move to the charge natural point with the increase of U, indicating a flatter band at a larger U.

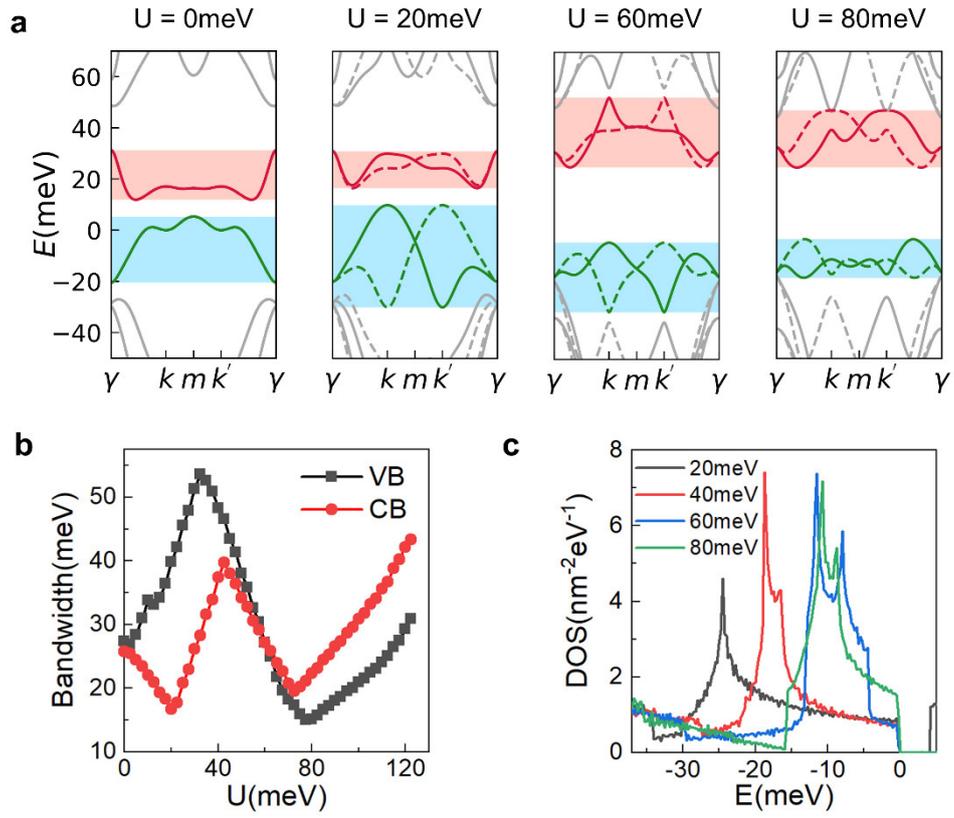

**Fig. S11. calculated band structures and density of states. a** Band structures at different interlayer potential energy with a twisted angle $\theta = 1.38°$. Red (blue) region corresponds to the first moiré conduction (valance) band. **b** Bandwidth of first moiré valance and conduction bands versus U. **c** Density of states of the first moiré valance band at different U.

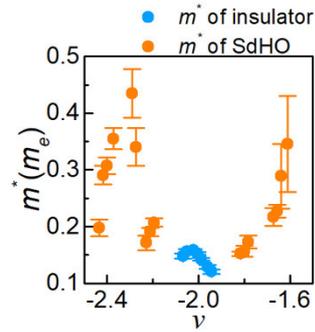

**Fig. S12. Effective mass.** Effective mass of SdHO and quantum oscillations of insulators corresponding to Fig. 1d of main text.

## Supplementary Note 5: Inverted band model

We calculate the Landau quantization in an inverted band model[5]. The Hamiltonian is

$$\begin{pmatrix} \frac{\hbar^2 k^2}{2m_K} - \mu_K & V\vec{k}\cdot\vec{\sigma} \\ V\vec{k}\cdot\vec{\sigma} & -\frac{\hbar^2 k^2}{2m_{K'}} - \mu_{K'} \end{pmatrix},$$

where $m_K = 0.15 m_e$ and $m_{K'} = m_K/\alpha$ are effective mass of K and K' valley subbands, respectively. $\alpha = 0.8$ corresponds to the heavier hole-like subband in our model. $\delta\mu = \mu_K - \mu_{K'} = \beta(D + D_b)$, determined by the electric displacement field $D$, shows the size of overlap of two bands. Here $\beta \approx 93.53*10^{-22}$ C·nm, $D_b \approx 1.26$V/nm are extracted from experimental results. $V$ corresponds to the size of hybridization gap. $\sigma = (\sigma_x, \sigma_y)$ are Pauli matrices. We ignore the dynamic change of band overlap with the increase of $B$ in this model because of the small Zeeman split energy relative to the large size of band overlap, which is reasonable in optimal oscillation region of $D$ from -0.7 to -1.1 V/nm.

We calculate the low energy density of states at finite temperature $T = 2$K using a formula $D_T = \int_{-\infty}^{+\infty} \frac{\partial n_F(E-\mu,T)}{\partial \mu} D(E) dE$, where $n_F$ is Fermi-Dirac distribution, $\mu$ is the chemical potential in gap. $D_T$ can be expressed as $\frac{eB}{2\pi\hbar} \frac{1}{2T} \sum_{n=0}^{\infty} \frac{1}{\cosh(\frac{E_n-\mu}{T})+1}$ in magnetic field. Here $E_n$ is Landau level spectrum.

Fig. S13 shows the Landau fan diagram in experiments (a) and calculations (b). The inverted band model reproduces the intersection of Landau levels due to the inverted multiple Fermi surfaces near $v = -2$. These intersecting patterns provide solid evidence of inverted band in twisted double bilayer graphene.

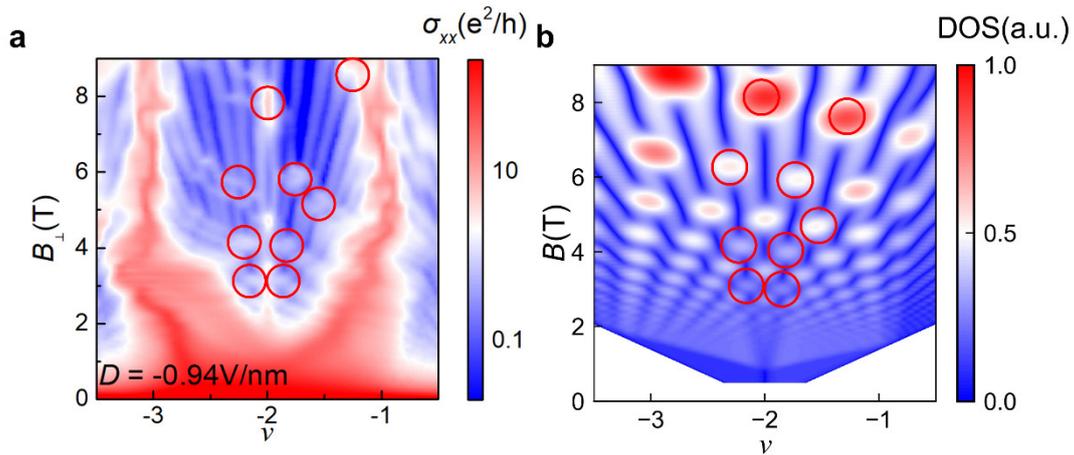

**Fig. S13. A comparison of experimental LLs spectra near $v = -2$ and the ones calculated from inverted band model. a** $\sigma_{xx}$ as a function of $v$ and $B_\perp$ at $D = -0.94$V/nm. **b** Calculated LLs diagram according to inverted band model. The DOS in red circles are higher than other regions, corresponding to the intersection of LLs emanating from two inverted bands.

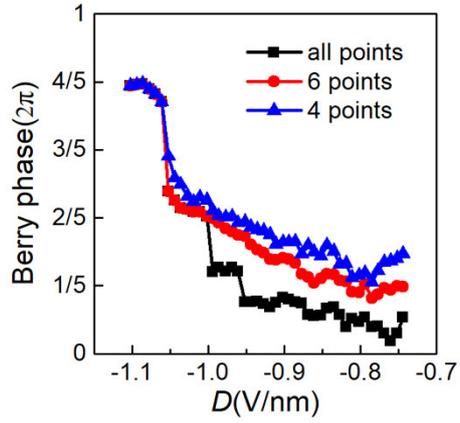

**Fig. S14. Berry phase extracted from quantum oscillations index at $v = -2$.** We calculated berry phase by fitting different numbers of oscillation peak and valley due to $D$ dependence of numbers of index. There are at least four indices in the whole range of $D$.

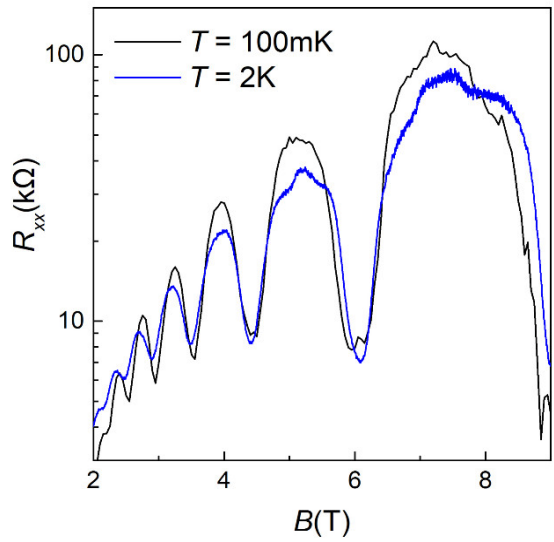

**Fig. S15. Significant QOs at ultra-low temperature in dilution refrigerator.**

**Supplementary reference**